\documentclass[11pt]{article}
\usepackage{hyperref}
\pdfoutput=1
\begin{document}

\title{Deconstructing wall turbulence -\\visualization of resolvent modes}%
\author{Daniel Barella$^1$, Sarah Churng$^2$, Conrad Egan$^3$,\\
 Rashad Moarref$^4$, Mitul Luhar$^4$, Hillary Mushkin$^4$,\\ Scott Davidoff$^5$, Maggie Hendrie$^6$ \& Beverley J. McKeon$^4$\\
\\
%\vspace{6pt}
$^1$ Oberlin College\\
$^2$ University of Washington\\
$^3$ Texas A\&M University\\
$^4$ California Institute of Technology\\
$^5$ NASA Jet Propulsion Laboratory, California Institute of Technology\\
$^6$ Art Center College of Art \& Design}

%\thanks{}%
%\date{September 30, 2013}%
% ----------------------------------------------------------------
\maketitle
\begin{abstract}

This article accompanies a fluid dynamics video entered into the Gallery of Fluid Motion of the 66th Annual Meeting of the APS Division of Fluid Dynamics.
\end{abstract}

% ----------------------------------------------------------------
\section{Introduction}

In recent work  we have demonstrated that key features of wall turbulence can be captured by an input-output relationship between nonlinear forcing and velocity response, where the transfer function, constructed in wavenumber-frequency space, is named the \textit{resolvent}\cite{McKeon2010, McKeonPoF13, Sharma13}. A basis for the wall-normal direction can be obtained by singular value decomposition of the resolvent, where the singular functions (or resolvent modes) represent the most amplified velocity response to the ``most dangerous'' input forcing. As such, a low-rank approximation can be obtained by retaining a limited number of singular functions; recognizable statistical and structural features of wall turbulence can be identified in even the rank-1 approximation, in which only the principal singular functions are investigated.

For fully-developed, statistically stationary channel flow, the output of this analysis at each $(k_x,k_z,\omega)$, where $k_x$ and $k_z$ are the streamwise and spanwise wavenumbers, respectively, and $\omega$ is the frequency, consists of a three-dimensional, three velocity component propagating wave which we hypothesize can be considered to represent a ``building block'' of wall turbulence.  The three-dimensional and propagating nature of these resolvent modes presents a significant challenge for visualizing individual building blocks as well as the complexity associated with assemblies of even limited numbers of resolvent modes.

In the linked videos, we report on the outcome of a design-informed visualization project performed by the Caltech/JPL/Art Center Data Visualization Summer Internship Program tasked with creating a visualization of the deconstruction of a full turbulent flow field into subsets of dynamic, interacting building blocks. The goal was to create a visualization suite capable of illuminating the velocity fields associated with individual resolvent modes and the rapid increase in complexity associated with linear superposition of modes. The software, named \textit{ModeDynamix} has the potential to provide insight into the underlying dynamics of wall turbulence.

\section{Videos}

%\href{URL of video}{name that will appear in the text}
High and low resolution versions, respectively, of the video entry to the 2013 Gallery of Fluid Motion are shown in\\ \\
\vspace{12pt}
\textit{V102376} and\\
\vspace{6pt}
\textit{V102376SMALL}. \\

In these movies, the following phenomena are shown:

\begin{itemize}
\item A brief schematic introduction to resolvent analysis and description of the turbulence kernel, consisting of three resolvent modes, that is explored in the video;
\item Visualization of the velocity field and vortical structure associated with single resolvent modes, achieved using a combination of vector plots and tracking of massless particles released in the flow. Structures observed include quasi-streamwise vortices, vortical arches, inactive ``sloshing'' in the wall-parallel plane and meandering streaks;
\item Visualization of the complex velocity field arising from the superposition of only the three resolvent modes constituting the kernel;
\item A demonstration of the motion experienced by an infinitesimal, massless particle in the turbulence kernel.
\end{itemize}

\section*{Acknowledgements}

The visualization software was produced as part of the Caltech/JPL/Art Center Data Visualization Summer Internship Program. This research was carried out at the Jet Propulsion Laboratory, California Institute of Technology, under a contract with the National Aeronautics and Space Administration and at the California Institute of Technology with support from the Air Force Office of Scientific Research, grants FA-9550-09-1-0701 and FA-9550-12-1-0469. Copyright 2013 California Institute of Technology.

\end{document}